\documentclass{article}
\usepackage{amsmath,amssymb,latexsym,amsthm,amscd,amsxtra}
\usepackage[all]{xy}
\usepackage{epsfig}
\usepackage{graphicx}
\usepackage{caption}
\usepackage{subcaption}
\usepackage{epstopdf}
\usepackage{cite}
\usepackage[bottom]{footmisc}
\usepackage[section]{placeins}
\usepackage{color}

\graphicspath{{./images/}}

\begin{document}

\begin{center}
{\huge{\bf Magnet traveling through a conducting pipe: a variation on the  analytical approach}}\\
\vspace*{.5 in}
{\large Benjamin Irvine\footnote{e-mail: birvine@luc.edu}, Matthew Kemnetz\footnote{e-mail: mkemnetz@luc.edu}
Asim Gangopadhyaya\footnote{e-mail: agangop@luc.edu}, and Thomas Ruubel\footnote{e-mail: truubel@luc.edu}
}

\vspace*{.1 in}
{\it Department of Physics, Loyola University Chicago, Chicago, Illinois 60626}
\end{center}

\vspace*{.35 in}
\begin{abstract}
We present an analytical study of magnetic damping. In particular, we investigate the dynamics of a cylindrical neodymium magnet as it moves through a conducting tube. Owing to the very high degree of uniformity of the magnetization for neodymium magnets, we are able to provide completely analytical results for the EMF generated in the pipe, and the consequent retarding force. Our analytical expressions are shown to have excellent agreement with experimental observations.
\end{abstract}

\vspace*{.1 in}
\noindent
PACS: 41.20.Gz; 75.50.Ww; 75.50.Dd\\
Keywords: Faraday's Law, Electromagnetic Damping, Regenerative Braking


\section{Introduction}
\label{sec:introduction}

Magnetic braking plays a significant role in industry.  It is used to slow down the moving parts of systems without losing energy to friction.  In addition, the absence of frictional forces and direct physical contact between moving parts helps these parts last longer.  Thus, an improved understanding of magnetic damping is important to the development of future technology in regenerative braking.  In industry, complex computational models are often used to simulate realistic scenarios of magnetic braking.  We have developed a fully theoretical model for a cylindrically symmetric system, which can be used to benchmark these computational models.

We present here an analysis of a common demonstration that comprises  a cylindrical magnet and a non-ferromagnetic conducting tube in relative motion to each other.
\cite{Wiederick-1987,Heald-1988,Clack-1990,Levin-2006,Iniguez2-2007,Iniguez3-2007,Ireson-2008,
Marcuso1-1991,Marcuso2-1991,MacLatchy-1993,
McCarthy1996,Aguirregabiria-1997,Iniguez1-2004,Partovi-2006,Knyazev-2006,Bae-2009,Donoso1-2009,
Donoso2-2009,Derby2010,Salzman-2001,Hahn-1998}. Owing to the interaction between the moving magnet and the induced current in the pipe, the magnet falls very slowly through the tube, always generating a sense of amazement in students and teachers alike. This area has been explored by many researchers \cite{Wiederick-1987,Heald-1988,Levin-2006,Iniguez2-2007,Marcuso1-1991,Marcuso2-1991,MacLatchy-1993,
McCarthy1996,Aguirregabiria-1997,Iniguez1-2004,Partovi-2006,Knyazev-2006,Bae-2009,Donoso1-2009,
Donoso2-2009,Derby2010,Salzman-2001,Hahn-1998}.

In this paper we study the motion of a cylindrical neodymium magnet through a copper pipe of circular cross-section. The azimuthal symmetry of the problem keeps the mathematics tractable and allows us to generate an analytical expression for the EMF generated in an arbitrary segment of the tube, and the resulting retarding force.

Our paper is organized as follows. In Sec. \ref{sec:experimental setup}, we will describe the experimental setup used for this demonstration.  In Sec. \ref{sec:magnetic field}, we develop our model assuming the near-uniformity of magnetization of neodymium magnets, and then show that the resulting prediction of the magnetic field strength has excellent agreement with the measured values of the field on the axis of the magnet. We also compare the experimental results with the often used point dipole approximation. In Sec. \ref{sec:comp of flux}, from the model constructed in the previous section, we compute the flux through circular loops of the conducting pipe and generate an expression for the  current in a section of pipe of arbitrary length. As a special case, in Sec. V, we also compute the current generated in the forward half of the pipe (or alternatively in the wake of the magnet). In Sec. \ref{sec:comp of retarding force}, we compute the force on the magnet due to the interaction between the magnet and the pipe. Our analytical results match extremely well with experimental observations.
In the next section, we describe our experimental setup.

\section{Experimental Setup}
\label{sec:experimental setup}
\begin{figure}[htb]
   \centering
        \begin{subfigure}[t]{.35\textwidth}
               \centering
                \includegraphics[width=0.65\textwidth]{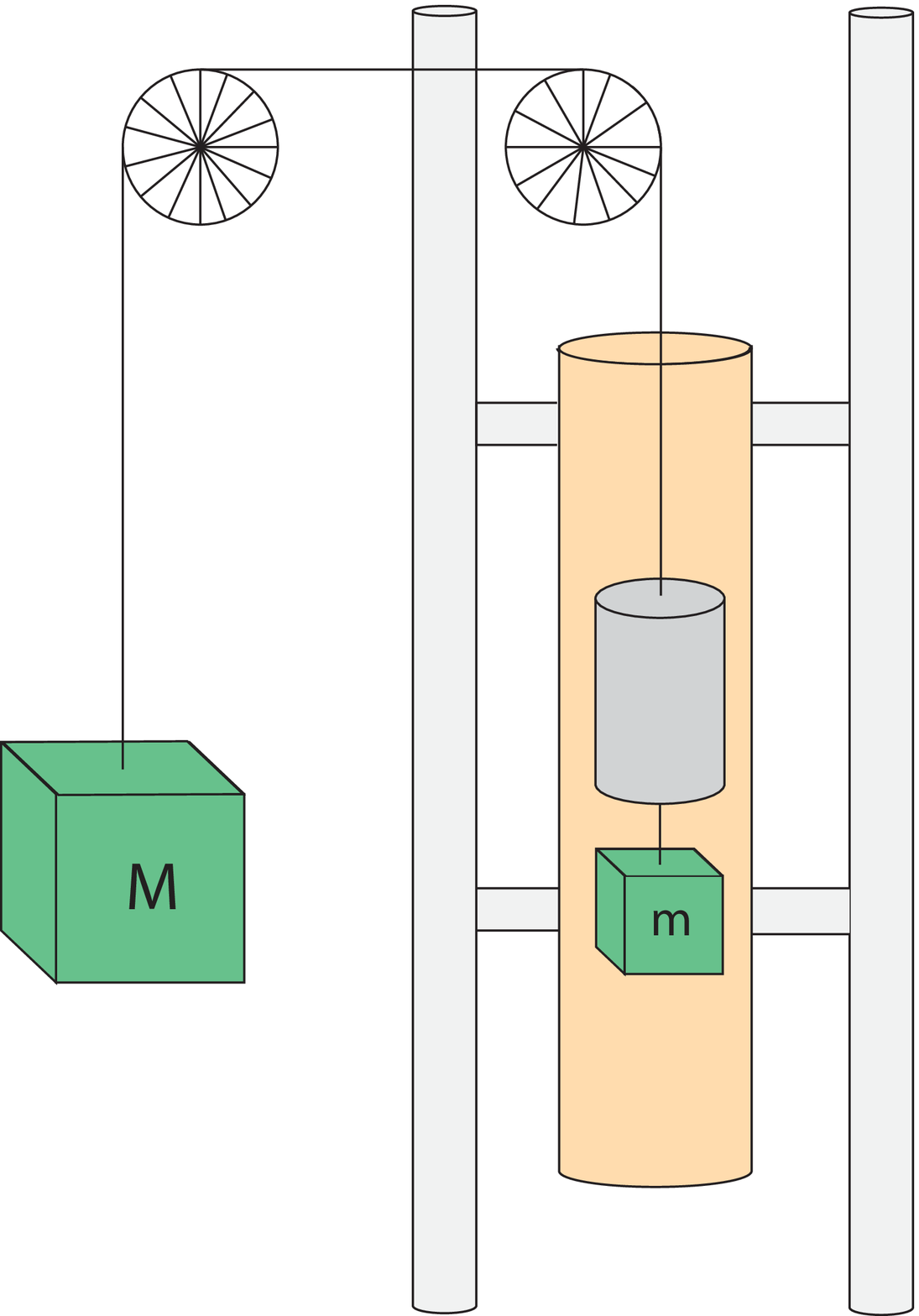}
                \caption{Experimental setup of copper pipe and neodymium magnet attached to pulley system.}
                \label{fig:velocity setup}
      \end{subfigure}
\quad\quad\quad
        \begin{subfigure}[t]{.35\textwidth}
                \centering
                \includegraphics[width=1.250\textwidth]{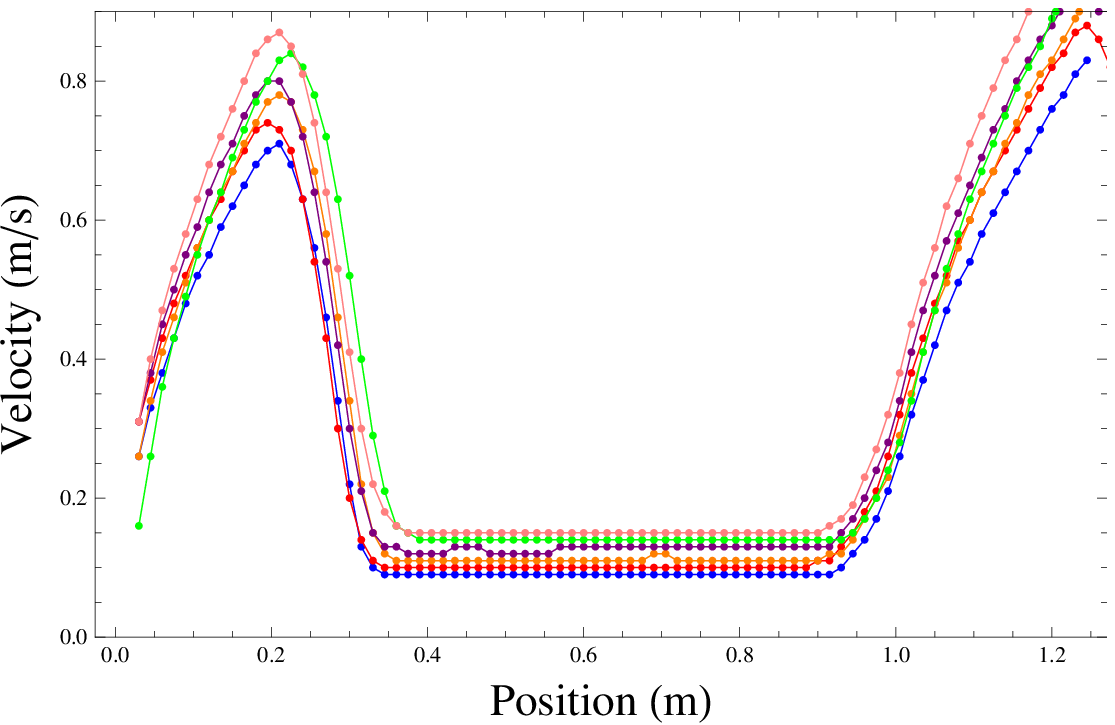}
                \caption{Velocity of magnet as it enters and travels through the conducting pipe.}
                \label{fig:velocity and acc runs}
        \end{subfigure}
        \caption{}\label{fig:freshman plots}
\end{figure}

As shown  in Fig. (\ref{fig:velocity setup}), we used hanging masses, $m$ and $M$, to pull a  cylindrical neodymium magnet through a copper pipe with varying terminal velocities. We used smart pulleys from PASCO to record the position, velocity, and acceleration of the magnet as it traveled into, through, and out of the pipe. Fig.  (\ref{fig:velocity and acc runs}) shows that for a significant segment of each individual trajectory of the magnet, the velocity remains constant.

We also find that the dependence of the resistive force on the terminal velocity can be accurately modeled by a linear relation.  As we show in Sec. \ref{sec:comp of retarding force}, this linear behavior is replicated by our theoretical analysis as well.
Researchers have studied the damped oscillatory motion of a magnet in a conducting tube \cite{Hahn-1998}. However, in this work, we have limited ourselves to an analytical study of the emf and the retarding force for a magnet moving with different terminal velocities.

\section{Magnetic Field due to a Neodymium Magnet}
\label{sec:magnetic field}
In order to quantitatively express the magnetic field, we need to develop an appropriate model  of our magnet.  Several authors have considered the magnet to be a pure dipole \cite{{Iniguez2-2007},{Knyazev-2006},{Bae-2009},{Donoso1-2009},{Hahn-1998}}.  This model works well for small magnets moving through wide pipes. Some have also considered a physical dipole  constructed of two point monopoles separated by an appropriate distance \cite{Levin-2006}. This too would be a good approximation when the radius of the magnet is much smaller than the diameter of the pipe, and the monopoles are well inside the magnet; i.e., not too close to the surface.  Our aim is to keep the analysis general and accessible to undergraduate students.  In particular, we specifically include the case where the dimension of the magnet is comparable to the  diameter of the pipe and generates strong braking. For such cases, as we will show in Fig. (\ref{fig:AxialData}), the dipole model does not accurately fit the data.

Neodymium magnets have a very uniform magnetization. This uniformity allows us to simulate the $\vec{B}$-field of the cylindrical magnet by two circular disks with uniform magnetic surface charge densities, $\sigma_{m} $ and $-\sigma_{m} $, where  $\sigma_{m} $ is proportional to the magnetization density $M_o$ of the magnet \cite{Jackson}.  The method of determining the $\vec{H}$-field is then identical to the case of finding the electric field due to two uniformly charged parallel disks of surface charge densities, $\sigma_e$ and $-\sigma_e$. In \cite{Levin-2006}, the authors had recognized that applicability of the two-disk model for this case, however, they later chose to approximate it by a physical dipole consisting of two monopoles.

\subsection{Magnetism in a polarizable medium}
\label{subsec:magnetism in a polarized medium}
The magnetic field due to a current density $\vec{J}$ is given by
\begin{eqnarray}
{\rm Ampere's~ Law:}~~~~\overrightarrow{\nabla}\times \vec{B}  = \mu_o \vec{J} ~~.
 \label{eq:Ampere}
\end{eqnarray}
$\vec{J}$ includes the  ``free-currents" $\vec{J_f}$ and the bound current density $\vec{J}_b  = \overrightarrow{\nabla}\times \vec{M}$, where  $\vec M$ is the magnetization density (magnetic moment per unit volume).  Thus,  in the presence of magnetization, we have
\begin{eqnarray}
\overrightarrow{\nabla}\times \vec{B}  &=& \mu_o \left(\vec{J}_f + \vec{J}_b \right)
=\mu_o \left(\vec{J}_f + \overrightarrow{\nabla}\times \vec{M}  \right)~~.
\label{eq:ModifiedAmpere}
\end{eqnarray}
For a permanent magnet; i.e.,  $\vec{J_f}=0$,  eq. (\ref{eq:ModifiedAmpere}) yields:
\begin{eqnarray}
 \overrightarrow{\nabla}\times {\left( {\vec{B} } -{\mu_o}\vec{M}  \right)}&=&  \overrightarrow{\nabla}\times {{\mu_o}\vec{H} } = 0~~.
\end{eqnarray}
Where we have defined the conservative field $\vec{H}$ such that
$ {\vec{B} }  =  {\mu_o}\left( \vec{M}  + \vec{H}\right).$
Since $ \overrightarrow{\nabla}\cdot {\vec{B} } = 0$, we have
\begin{eqnarray}
\overrightarrow{\nabla}\cdot {\vec{H}}  &=&    -\overrightarrow{\nabla}\cdot \vec{M}  \label{eq:Poisson4H}~~.
\end{eqnarray}
Comparing this equation with Gauss' law
$\overrightarrow{\nabla}\cdot {\vec{E}}  = \frac{\rho_e}{\epsilon_o} $,
we see that the $\vec{H}$-field  is generated by the source $\rho_m \equiv -\, \overrightarrow{\nabla}\cdot \vec{M} $ exactly in the same way as the electrostatic field  $\vec{E} $ is found from the electrical charge density ${\rho_e}$.

\subsection{Magnetic Scalar Potential due to a magnet with uniform density $M_o \, \hat e_z$}
\label{subsec:magnetc scalar potential}
Since $\vec{H}$ is a conservative field,   we can write it as a gradient of a scalar field. I.e.,
${\vec{H}}  =   -\overrightarrow{\nabla} \Psi_m.$
From Eq. (\ref{eq:Poisson4H}), we have
\begin{eqnarray}
\overrightarrow{\nabla}^2 \Psi_{m} ~=~ -{\rho_m}  = \overrightarrow{\nabla}\cdot \vec{M} ~.\label{eq:Poisson4H2}
\end{eqnarray}

\begin{figure}[htb]
  \centering
    \includegraphics[width=1.0\textwidth]{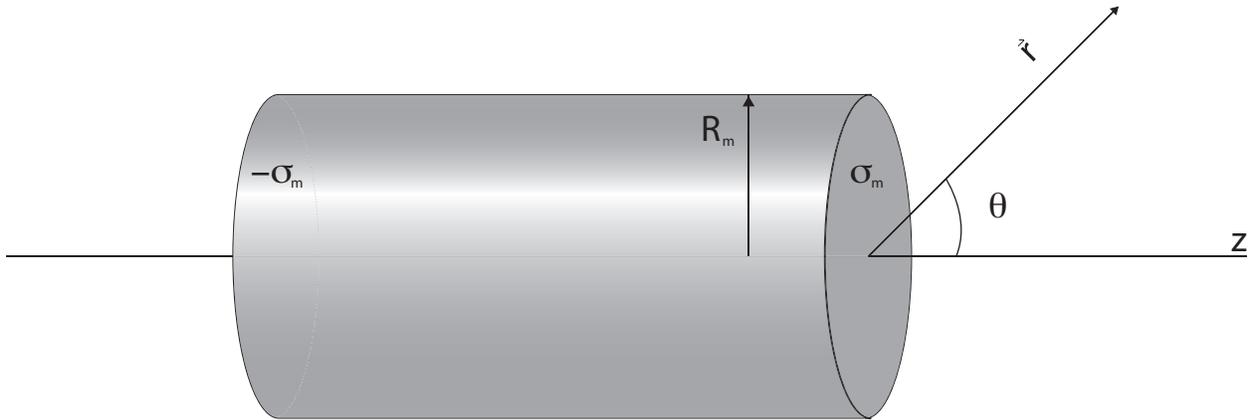}
  \caption{Two disks of uniform magnetic charge density $\pm \,\sigma_{m} $ are at $z=\pm \frac L2$ respectively.}
  \label{fig:magnetdiagram}
\end{figure}
For a cylindrical magnet with uniform magnetization density $M_o \, \hat e_z$, the $\overrightarrow{\nabla}\cdot \vec{M}$ is zero at all points inside the magnet, and receives non-zero contributions only at the two circular end surfaces. Hence, the $\vec H$-field generated by the cylindrical magnet is the same as that of two disks of uniform magnetic surface charge densities $\sigma_{m} $ and $-\sigma_{m} $ separated by a distance $L$, where $\sigma_{m}  =   M_o$. This expression for the $\vec H$-field would be  valid both inside and outside the magnet. The $\vec B$-field is then simply given by $\mu_o \vec H$ outside the magnet and $\mu_o \left(\vec H+\vec M\right)$ inside.

The $\vec H$-field on the axis of the magnet can be readily derived by superimposition of scalar potentials due to a single disk of uniform magnetic surface charge density $\sigma_m$:
\begin{eqnarray}
\Psi_m (z) &=& \frac{\sigma_m}{2} \left( z - \sqrt{R_m^2+ z ^2}   \right) =
\frac{M_o}{2} \left( z - \sqrt{R_m^2+ z ^2}   \right)~~.
\label{eq:1Disk-AxialField}
\end{eqnarray}
The scalar potential due to the cylindrical magnet is then given by \footnote{The expression derived in Eq. (\ref{eq:2Disk-AxialField}) assumes that the origin is set at the center of the magnet.}
\begin{eqnarray}
\Psi_m^{\rm 2-Disks} &=& \frac{M_o}{2}  \left[
\left( \left(z-\frac L2\right) - \sqrt{R_m^2+ \left(z-\frac L2\right) ^2}   \right)
-
\left( \left(z+\frac L2\right) - \sqrt{R_m^2+ \left(z+\frac L2\right) ^2}   \right)
\right]~.\label{eq:2Disk-AxialField}
\end{eqnarray}

\begin{figure}[ht]
  \centering
    \includegraphics[width=.85\textwidth]{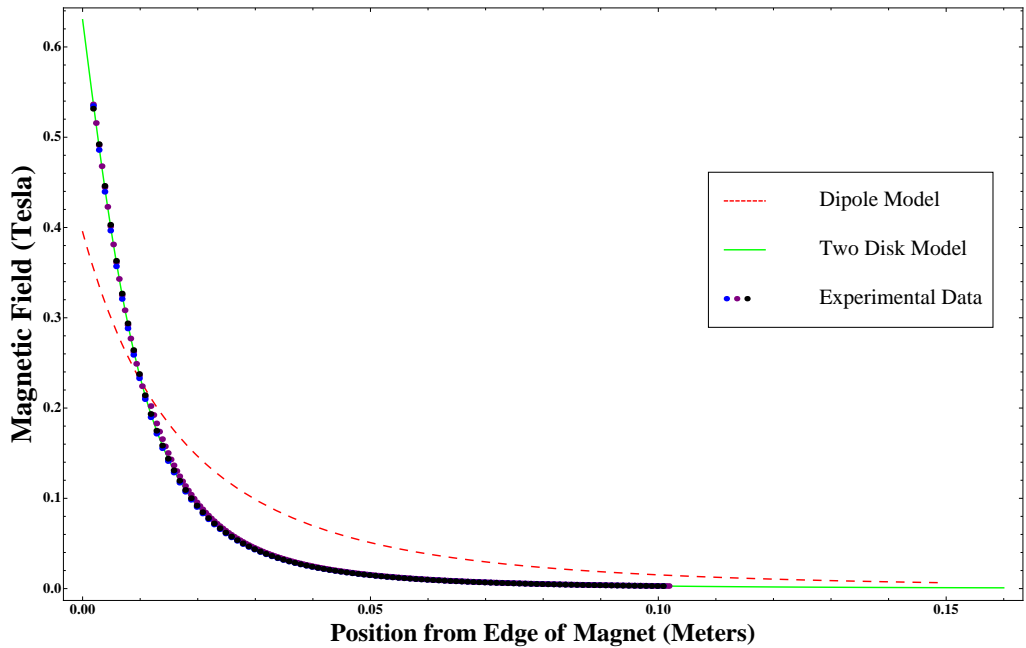}
  \caption{Axial magnetic field for the Dipole model (dashed), Two-Disk model (solid line) and the experimental data }
  \label{fig:AxialData}
\end{figure}

In Fig. (\ref{fig:AxialData}), we show a plot of the experimentally determined magnetic field against the values obtained from Eq. (\ref{eq:2Disk-AxialField}).  For comparison, we also plot the field due to a pure dipole with the net dipole moment equal to the dipole  moment of the magnet $ \left( \pi R_m^2 \, M_o\right) $. As is evident from Fig. (\ref{fig:AxialData}), our experimental data is in excellent agreement with the predictions of the two-disk model, and hence verifies our assumption regarding the uniformity of the neodymium magnets. Henceforth, our theoretical analysis will assume that the magnetization is uniform.

\subsection{Computation of the Magnetic Field due to the Cylindrical \\ Neodymium Magnet}
\label{subsec:comp of mag-field}
To compute the off-axis $\vec B$-field, we will start with the axial field given in Eq. (\ref{eq:1Disk-AxialField}). Except for points on  one of the circular end surfaces of the magnet, the magnetic scalar potential $ \Psi_m$ satisfies $\overrightarrow{\nabla}^2 \Psi_{m} = - {\rho_m} = 0$. Hence, the general solution for $ \Psi_m$  due to one disk in spherical coordinates \footnote{For this azimuthally symmetric problem, we have set the origin of the coordinates at the center of the disk, and $z$-axis coincides with the axis of the magnet.} is
\begin{eqnarray}
 \Psi_m(r,\theta) = \sum_{\ell=0}^{\infty} \left(a_\ell r^{\ell} + \frac{b_{\ell}}{r^{\ell+1}}   \right) {\cal P}_{\ell} (\cos\theta)~~ \label{eq:General-Phi}
\end{eqnarray}
As we will later see, for the calculation of flux, we will only need to work in the region $r > R_m$ \footnote{$R_m$ is the radius of the magnetic disk; i.e., the same as the radius of the magnet.}, hence all $a_\ell=0$, and the scalar potential is reduced to
\begin{eqnarray}
 \Psi_m(r,\theta) = \sum_{\ell=0}^{\infty} \frac{b_{\ell}}{r^{\ell+1}}  ~ {\cal P}_{\ell} (\cos\theta)~~. \label{eq:laplace}
\end{eqnarray}
In order to determine the values for constants $b_{\ell}$ in Eq. (\ref{eq:laplace}), we note that the expression for $\Psi_{m} (r,0)$ must equal $\Psi_{m} (z)$ of Eq. (\ref{eq:1Disk-AxialField}) when $z$ is replaced by $r\cos 0^o = r$;  i.e.,
\begin{eqnarray}
 \sum_{\ell=0}^{\infty} \left(\frac{b_{\ell}}{r^{\ell+1}}  \right) = \frac{\sigma_{m} }{2}\left[\left(R_m^{2} + r^{2}\right)^{\frac{1}{2}} -r\right] ~,\label{eq:solvebsubl}
\end{eqnarray}
where we have used ${\cal P}_{\ell} (1) = 1$ for all $\ell$.
By comparing the powers of $r$ on both sides, we find that all $b_{2 \ell+1}$ are zero, and the even coefficients $b_{2 \ell}$ are given by
\begin{eqnarray}
 b_{2 \ell}= \left[\frac{\sigma_{m} \,R_m^{2 \ell +2}}{2\left(\ell+1\right)!}\right] \prod_{k=0}^{\ell}  \left(\frac{1}{2}-k\right)~.
\end{eqnarray}
Thus the magnetic scalar potential $\Psi_{m} (r,\theta)$ is given by
\begin{eqnarray}
 \Psi_m(r,\theta) = \sum_{\ell=0}^{\infty} \left[\frac{\sigma_{m} \,R_m^{2 \ell +2}}{2\left(\ell+1\right)!}\right]  ~~\frac{\prod_{k=0}^{\ell}  \left(\frac{1}{2}-k\right)}{r^{2\ell+1}} ~~   {\cal P}_{2\ell} (\cos\theta).~~ \label{eq:laplace2}
\end{eqnarray}
In terms of $\Psi_{m} (r,\theta)$, we can find the magnetic field, $\vec{B}$ outside of the magnet by
\begin{eqnarray}
\vec{B} = -\mu_o\overrightarrow{\nabla} \Psi_{m}  ~,\label{eq:magfield_Out}
\end{eqnarray}
and for inside the magnet, we will need to add an additional term:
\begin{eqnarray}
\vec{B} = -\mu_o \left(  \overrightarrow{\nabla} \Psi_{m} - \vec{M}\right) ~.\label{eq:magfield_In}
\end{eqnarray}
Thus, we have an exact expression for the magnetic field. The sum can be computed to any desired level of accuracy by including a sufficiently large number of terms. In Ref. \cite{Partovi-2006}, Partovi et al. had carried out a very comprehensive analysis for a uniformly magnetized cylinder as well. However, they considered the vector potential due to the moving magnet. Similarly, the authors of \cite{Derby2010} computed the magnetic field and the flux due to a cylindrical magnet and reduced it to the computation of elliptical integrals that could be done using {\it Mathematica}. We find that, due to the similarity with electrostatics, the scalar potential method is  much more accessible to undergraduate students. In addition, by choosing to keep an appropriate number of terms in the expansion given in Eq. (\ref{eq:laplace2}), students can compute the scalar potential to any desired level of accuracy.

In the next section, we will use the expression of Eq. (\ref{eq:laplace2}) to evaluate flux through a cross-section of the pipe, a distance $z$ from the face of the magnet.

\clearpage
\section{Computation of Flux}
\label{sec:comp of flux}

As the magnet travels through the copper pipe, the changing magnetic flux causes eddy currents to form in the pipe.  We will assume that the pipe thickness is small compared to the radius of the pipe.  The authors of Refs. \cite{Knyazev-2006, Donoso1-2009, Partovi-2006} have studied the effect of thickness more carefully. We also assume that the magnet falls coaxially through the conducting pipe, and thus an azimuthal symmetry is maintained throughout the motion. In this case, the eddy currents generated in the pipe would form perfect circles perpendicular to the axis of symmetry. We will now carry out surface integrations of the magnetic field given by Eqs. (\ref{eq:magfield_Out}) or (\ref{eq:magfield_In}) to determine the flux through a circular cross-section of the pipe.  However, instead of computing the flux on a planar surface through the circle, we choose a spherical surface that contains the circle, and is centered at the center of the front-disk of the magnet.
\begin{figure}[ht]
  \centering
    \includegraphics[width=.85\textwidth]{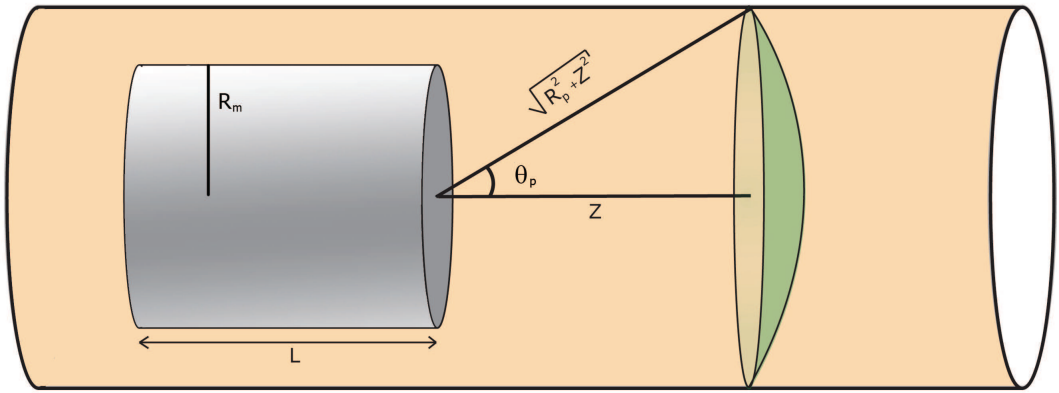}
  \caption{Pipe diagram}
  \label{fig:pipediagram}
\end{figure}
The flux $\Phi_{m} (z)$ through a circular loop at a distance $z$ from the front-disk is then given by
\begin{eqnarray}
    \Phi_{m} (z) &=& \int_S \! \vec{B} \cdot \hat{r} \, \mathrm{d} a = -\mu_o\int_S \! \frac{\partial \Psi_{m} (r,\theta)}{\partial r} da \nonumber\\\nonumber\\
&=&\sum_{\ell=0}^{\infty} 
  ~~b_{2\ell} \left(\left. \frac{\partial}{\partial r}\frac1{r^{2\ell+1}}\right|_{r=\sqrt{R_p^2+z^2}}\right) ~\int_0^{\theta_p}   {\cal P}_{2\ell} (\cos\theta)~\sin\theta d\theta\,d\phi~
\nonumber\\\nonumber\\
    &=& 2 \pi\mu_o \sum_{\ell=0}^{N} \prod_{k=0}^{\ell}  \left(\frac{1}{2}-k\right)\left[\frac{\sigma_{m} \,R^{2 \ell +2}}{2\left(\ell+1\right)!}\right]  \frac{(2\ell+1)}{(R_p^2+z^2)^\ell}\int^1_{u_p}
 \mathcal{P}_{2\ell}\left(u\right)d(u) ~,
\label{eq:fluxintegral}
\end{eqnarray}

\vspace*{0.1in}\noindent
where we have substituted $u=\cos\theta$, $u_p=\frac{z}{\sqrt{R_p^2+z^2}}$, $b_{2\ell} = \left[\frac{\sigma_{m} \,R_m^{2 \ell +2}}{2\left(\ell+1\right)!}\right]  ~\prod_{k=0}^{\ell}  \left(\frac{1}{2}-k\right) $, and have used $b_{2\ell+1}=0$ for all $\ell$.
We can compute this integral using the identity
$\mathcal{P}_{2\ell}\left(u\right)=\frac{1}{4\ell+1}\left(\frac{d P_{2\ell+1}}{du} - \frac{d P_{2\ell-1}}{du}\right)$
and get
\begin{eqnarray}
\Phi_{m} (z)
&=&  \frac{2\pi\mu_o\sigma_{m} }{2\sqrt{R_p^2+z^2}}\left[\sqrt{R_p^2+z^2}-z\right]
+ 2\pi\mu_o\sum_{\ell=1}^{N}\frac{(2\ell+1)b_{2\ell}} {(R_p^2+z^2)^\ell}\left[\frac{1}{4\ell+1}\right]
\nonumber\\ &\times& \left[\mathcal{P}_{2\ell-1}\left(\frac{z}{\sqrt{R_p^2+z^2}}\right)- \mathcal{P}_{2\ell+1}\left(\frac{z}{\sqrt{R_p^2+z^2}}\right)\right].
\end{eqnarray}

\begin{figure}[htb]
        \centering
        \begin{subfigure}[t]{0.4\textwidth}
                \centering
                \includegraphics[width=\textwidth]{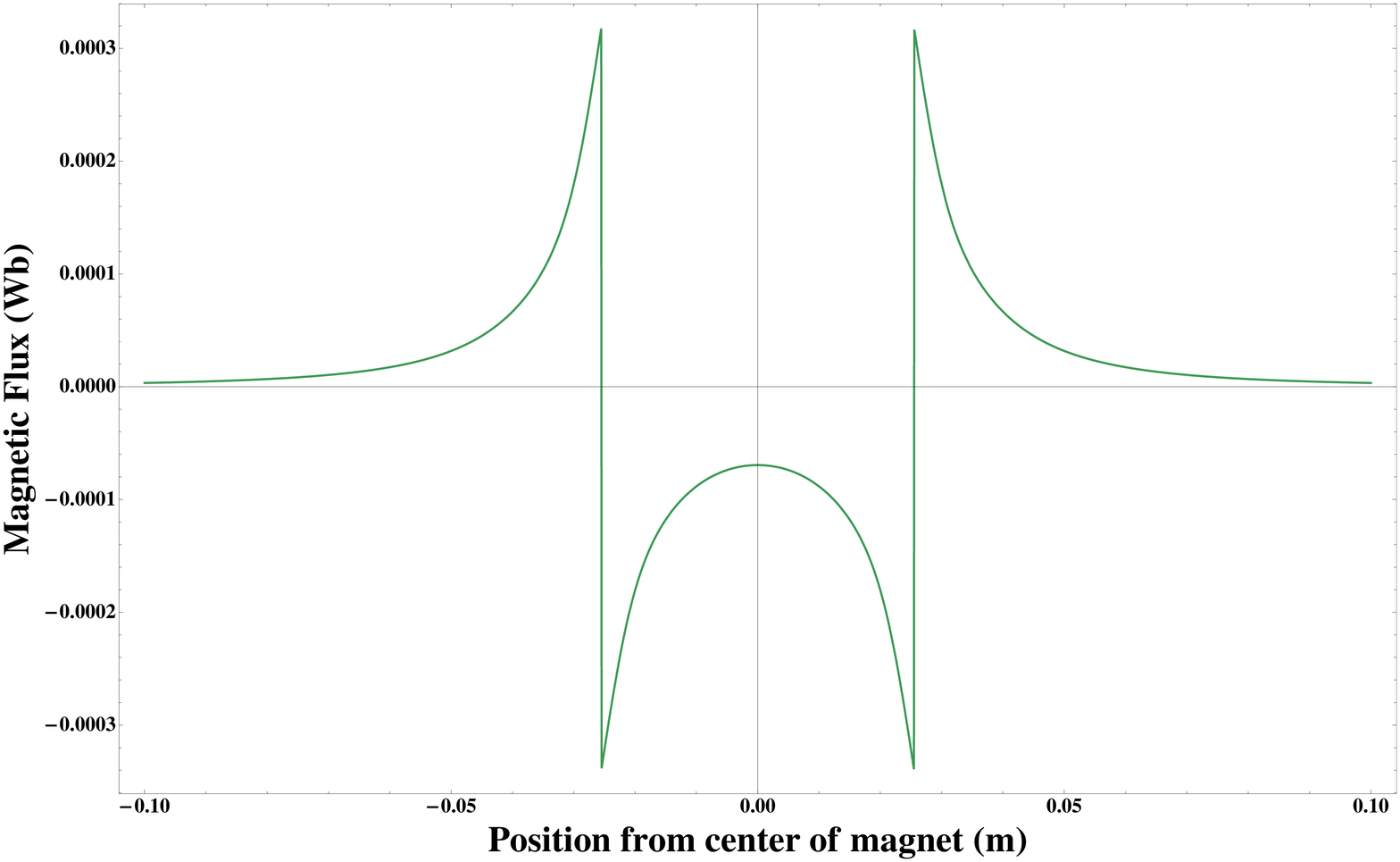}
                \caption{Contribution of the $\vec H$-field toward flux  $\Phi_{m} (z)$. This part of the flux has a discontinuity across each face of the magnet.}
                \label{fig:H}
        \end{subfigure}
\quad\quad%
        ~ 
        \begin{subfigure}[t]{0.4\textwidth}
                \centering
                \includegraphics[width=\textwidth]{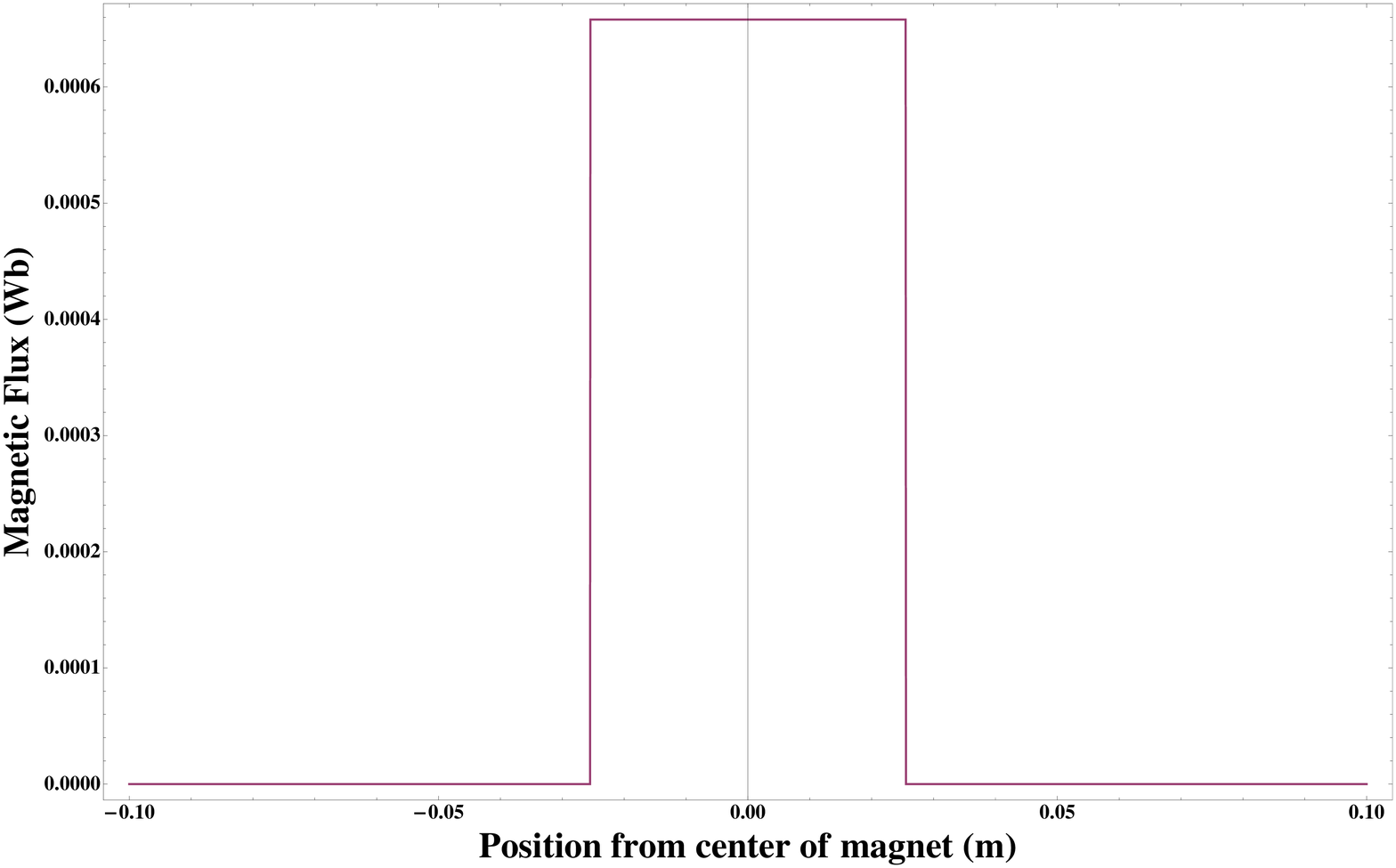}
                \caption{Contribution of the magnetization $\vec M$. }
                \label{fig:M}
        \end{subfigure}
\quad
        \begin{subfigure}[htb]{0.45\textwidth}
                \includegraphics[width=\textwidth]{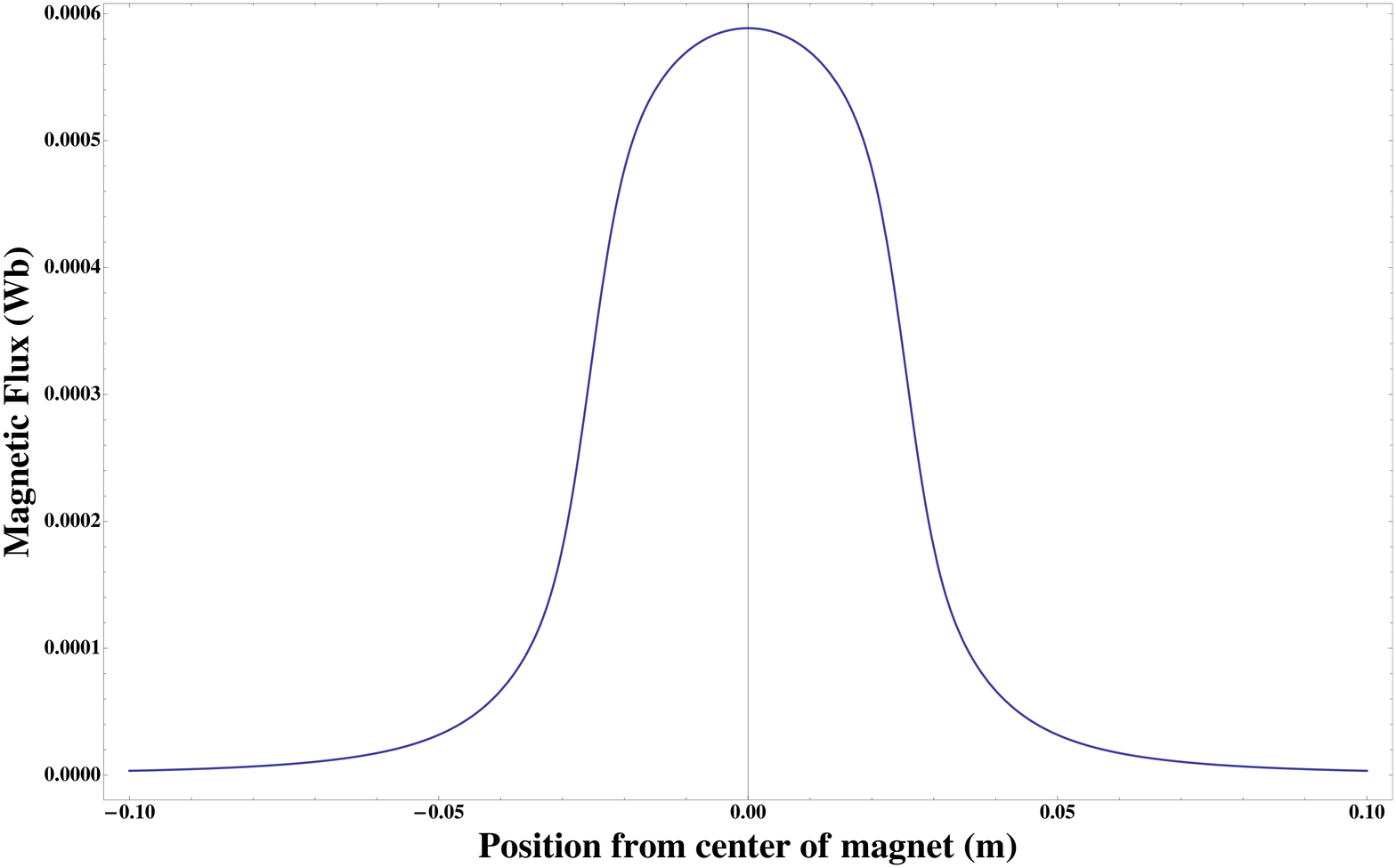}
                \caption{Magnetic flux $\Phi_m(z)$ given by $\int_z \vec B \cdot d\vec a$.}
                \label{fig:B}
        \end{subfigure}
        \caption{Contributions of the various fields toward flux  $\Phi_{m} (z)$ through a cross-section of the pipe at a distance $z$ from the center of the magnet.}\label{fig:animals}
\end{figure}

\vspace*{0.1in}\noindent
Please note that the above expression for $\Phi_{m} (z)$  gives the flux due to one disk, measured from the center of that disk. To compute the flux due to the magnet, we need to consider two disks with magnetic charge densities $\sigma_m$ and $-\sigma_m$ separated by a distance $L$. The net flux is then given by the summation of the contributions from two disks situated at two planar faces of the magnet. In Figs. (\ref{fig:H},\ref{fig:M},\ref{fig:B}), we have plotted the contributions of the $\vec H$-, $\vec M$- and $\vec B$-fields  toward flux  $\Phi_{m} (z)$ through a circular cross-section of the pipe situated at a distance $z$ from the center of the pipe.. As expected, a superposition of Figs. (\ref{fig:H},\ref{fig:M}) generates the Fig. (\ref{fig:B}).
\vspace*{0.3in}\noindent

%

\section{Computation of EMF}
\label{sec:comp of emf}


Assuming the magnet to be moving with a constant velocity $v_o \hat z$, in this Sec., we will determine the time variation of the flux $\Phi_{m}$ through a loop as the magnet comes towards it, and then  passes through it. 

In order to compute the emf through a circular cross-section of the conducting pipe at a distance $z$ from the center of the magnet, we need to determine the change in flux through the loop during a time interval $\Delta t$.  During this time interval the distance of the loop from the magnet changes from $(z-\Delta z)$ to $z$. Hence, the change in flux $ \Delta \Phi_{m} $ seen by a loop is: $\Phi_{m} (z-\Delta z)-\Phi_{m} (z) = - \Delta z\, \left[\frac{\partial\Phi_{m} (z)}{\partial z}\right]$.  Since, this change happens during the time $\Delta t$ in which the magnet moves a distance $\Delta z = v_o\,\Delta t$, the emf is given by
\begin{eqnarray}
{\cal E} = \oint \vec E\cdot d\vec \ell = -\,\frac{\Delta \Phi_{m}}{\Delta t} = -\frac{ - \Delta z\, \frac{\partial\Phi_{m} (z)}{\partial z}}{\Delta t} = v_o \cdot \frac{\partial\Phi_{m} (z)}{\partial z}~.
\end{eqnarray}
The electric field in the wall of the pipe is then given by $E_{\phi} = \frac{v_o}{2\pi R_p}\, \cdot \frac{\partial\Phi_{m} (z)}{\partial z}$, and hence the current density in the pipe will be given by $J_{\phi} = \sigma_c\,E_{\phi} = \frac{v_o\,\sigma_c}{2\pi R_p}\, \cdot \frac{\partial \Phi_{m} (z)}{\partial z}$. Here, $R_p$ denotes the average radius of the pipe. The current $dI$ through a section of the pipe of thickness $\delta$ and length $dz$ will be given by
\begin{eqnarray}
 dI  = J_{\phi} \,\delta \, dz = \frac{v_o\,\sigma_c\,\delta}{2\pi R_p}\, \cdot \frac{\partial \Phi_{m} (z)}{\partial z}~dz~~.\label{eq.DifferentialCurrent}
\end{eqnarray}
Hence, the total current through a section of the pipe from $z_1$ to $z_2$ is then given by
\begin{eqnarray}
I = \frac{v_o\,\sigma_c\,\delta}{2\pi R_p}\, \left(\Phi_{m} (z_2) - \Phi_{m} (z_1)\right). \label{eq:DeltaCurrent}
\end{eqnarray}
\begin{figure}[htb]
    \centering
        \begin{subfigure}[t]{.4\textwidth}
                \centering
                \includegraphics[width=.5\textwidth]{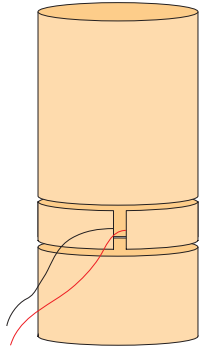}
                \caption{Setup for measuring current.}
                \label{fig:current setup}
        \end{subfigure}
~\quad\quad\quad
       \begin{subfigure}[t]{.5\textwidth}
                \centering
                \includegraphics[width=1\textwidth]{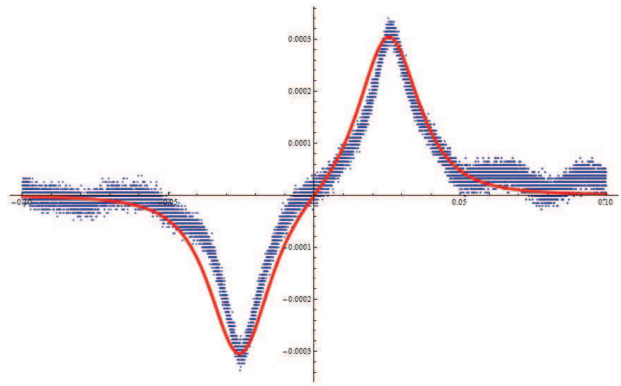}
                \caption{Plot of experimentally observed current and predicted current (solid line).}
                \label{fig:Measuredandmodeledcurrent}
        \end{subfigure}
        \caption{Experimental Setup}
\end{figure}

In order to verify the above expression for the current $I$, we took a small cylindrical slice from the middle of the pipe. We then cut a vertical slit down the spine of the above slice and replaced it between two longer segments of the pipe, as shown in Fig. (\ref{fig:current setup}).  We then wired the slice to an ammeter that recorded the current generated as a function of time \footnote{We actually used the MyDAQ device made by National Instruments to observe the generated current.}. Fig. (\ref{fig:Measuredandmodeledcurrent}) shows the current generated in a loop as the magnet passes through it. The solid line, in the background of the experimental data points collected by the MyDAQ, represents the current predicted by our model.  Please note that while the general behavior of the solid line
is given by Eq. (\ref{eq:DeltaCurrent}), the constants needed for the graph \footnote{The horizontal and vertical ranges of the graph were determined by requiring that  the crest and the trough of the theoretical graph match with the experimental data.} were obtained by stipulating that two points of the graph, namely the maximum and the minimum, matched with the corresponding points of the experimentally obtained data set.

In a long pipe, the total current in the part of the pipe that the magnet is yet to travel through, is given by
\begin{eqnarray}
I = \frac{v_o\,\sigma_c\,\delta}{2\pi R_p}\, \left[\Phi_{m} (\infty) - \Phi_{m} \left(0\right)\right].
\end{eqnarray}
In the next section, we will use Eq. (\ref{eq.DifferentialCurrent}) to compute the energy loss through a circular section of pipe of thickness $dz$, and from it the energy lost through an arbitrary segment of the pipe.
\section{Computation of Retarding Force}
\label{sec:comp of retarding force}
Since the magnet travels with a constant velocity, conservation of energy stipulates that the thermal loss in the conducting pipe per unit time will be equal to $\vec v\cdot\vec F$.  Thus, if we know the power loss, we  will be able to determine the force from the power loss. To compute the power loss in the pipe, we first determine the differential loss over an infinitesimal length $\Delta z$ of the pipe. This loss will be given by
\begin{eqnarray}
 dP &=& (dI)^2 (dR)
      ={\left(J_{\phi} \,\delta \, \Delta z\right)^2}\,\times\,{{\rm Resistance~ of~ length} ~dz} \nonumber\\\nonumber\\
    &=& \left( \frac{v_o\,\sigma_c\,\delta}{2\pi R_p}\, \cdot \frac{\partial \Phi_{m} (z)}{\partial z}~\Delta z \right)^2\cdot
 \frac{2\pi R_p}{\sigma_c\,\delta\,\Delta z}~,\nonumber\\\nonumber\\
 &=& v_o^2~ \frac{\sigma_c\,\delta}{2\pi R_p}\, \cdot \left( \frac{\partial \Phi_{m} (z)}{\partial z} \right)^2\cdot \Delta z~ .
\end{eqnarray}
Hence the total power loss is given by
\begin{eqnarray}
 P =  v_o^2~ \frac{\sigma_c\,\delta}{2\pi R_p}\, \cdot \int_{-\infty}^\infty \left( \frac{d\Phi_{m} (z)}{dz} \right)^2\cdot dz~ =2~v_o^2~ \frac{\sigma_c\,\delta}{2\pi R_p}\, \cdot \int_{0}^\infty \left( \frac{d\Phi_{m} (z)}{dz} \right)^2\cdot dz~.
\end{eqnarray}
The retarding force $F$ can then be derived using $P=\vec F\cdot \vec v = v_o\, F$. Thus, the force $F$ is given by
\begin{eqnarray}
F =  2~ \frac{v_o\,\sigma_c\,\delta}{2\pi R_p}\, \cdot \int_{0}^\infty \left( \frac{d\Phi_{m} (z)}{dz} \right)^2\cdot dz~.
\end{eqnarray}

\begin{figure}[htb]
   \centering
                \centering
                \includegraphics[width=0.6\textwidth]{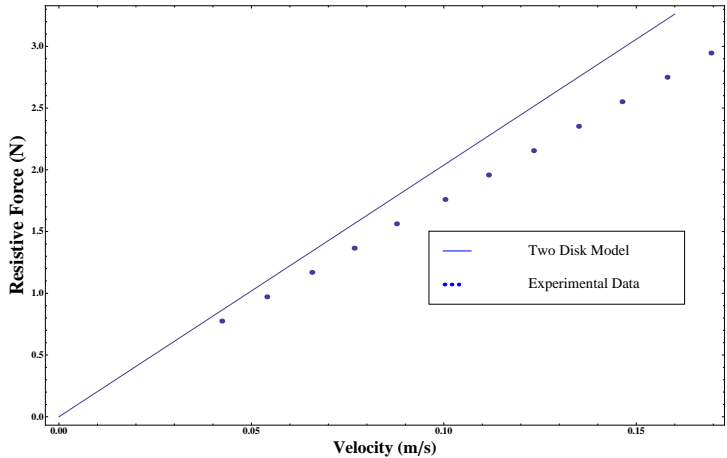}
                \caption{Experimentally oberved and theoretically computed (solid line) values of the resistive force for various terminal velocities.}
                \label{fig:resistive force vs run}
\end{figure}

Thus, we find that the resistive force is proportional to {\Large $\frac{v_o\,\sigma_c\,\delta}{R_p}$}.  In particular, if all other parameters are kept constant, we find $F\propto v_o$. Fig. (\ref{fig:resistive force vs run}) clearly exhibits this behavior in both experimental data as well as the theoretical model. It is important to point out that authors of Ref. \cite{ Partovi-2006} have shown that for speeds of less than 25 m/s, the linear-relation between the speed and the resistive force is an excellent model.

For computation, we chose to use the International Annealed Copper Standard (IACS) value of $5.8\ X\ 10^{7}\ S/m$ for $\sigma_c$ in our model because we were not certain of the specific alloy our copper pipe was made from. Recognizing that many commercially available copper pipes, like the one we used, have a conductivity closer to 90\% of the IACS, could explain why our predicted resistive force is slightly higher than what we observed experimentally.

\section{Conclusion}
\label{sec:ConcludingRemarks}
We studied the effect of a cylindrical neodymium magnet moving along the axis of a cylindrical conducting pipe. Using the symmetry of the setup and the excellent uniformity of the magnetization density of a neodymium magnet, we were able to develop an analytical model for the induced surface current density and resulting retarding force. The analytically predicted current distribution and the retarding force show excellent agreement with experimental observation. Since we used the scalar method that bears a close resemblance to electrostatics, our analysis is comparatively more accessible to  undergraduates.  In addition, students can compute the flux to a desired level of accuracy by keeping a sufficiently large number of terms in the expansion of the scalar potential.

For industrial applications, sophisticated computational models are used to understand the eddy currents, and the resulting magnetic braking. This analytical model could be used to verify the computational models.

\section{Acknowledgment} Two of the authors (BI and MK) would like to thank Loyola University Chicago for the Mulcahy scholarship, which helped make their undergraduate research possible. AG would like to thank the Center for Experiential Learning at Loyola University Chicago for an Engaged Learning Faculty Fellowship that provided partial support for his research.   We would also like to thank Mr. Christopher Kabat for his help in designing the experimental setups.

\end{document}